# Middle-Solving F4 to Compute Gröbner bases for Cryptanalysis over GF(2)

Heliang Huang, Wansu Bao*

Zhengzhou Information Science and Technology Institute, Zhengzhou 450000, China

## ABSTRACT

Algebraic cryptanalysis usually requires to recover the secret key by solving polynomial equations. Faugère's F4 is a well-known Gröbner bases algorithm to solve this problem. However, a serious drawback exists in the Gröbner bases based algebraic attacks, namely, any information won't be got if we couldn't work out the Gröbner bases of the polynomial equations system. In this paper, we in-depth research the F4 algorithm over GF(2). By using S-polynomials to replace critical pairs and computing the normal form of the productions with respect to the field equations in certain steps, many "redundant" reductors are avoided during the computation process of the F4 algorithm. By slightly modifying the logic of F4 algorithm, we solve the univariate polynomials appeared in the algorithm and then back-substitute the values of the solved variables at each iteration of the algorithm. We call our improvements Middle-Solving F4. The heuristic strategy of Middle-Solving overcomes the drawback of algebraic attacks and well suits algebraic attacks. It has never been applied to the Gröbner bases algorithm before. Experiments to some Hidden Field Equation instances and some classical benchmarks (Cyclic 6, Gonnet83) show that Middle-Solving F4 is faster and uses less memory than Faugère's F4.

## KEYWORDS

Gröbner bases, F4, algebraic attacks, normal form, field equations, Middle-Solving F4, multivariate polynomial systems of equations

*Correspondence

Wansu Bao, Zhengzhou Information Science and Technology Institute, Zhengzhou 450000, China

E-mail: glhhl0773@126.com

## 1 INTRODUCTION

As one of the most efficient attacks, algebraic attacks have been successful in breaking several stream ciphers, public key cryptosystems, and a few block ciphers.

Algebraic attacks try to reformulate a cipher as a (very large) system of polynomial equations and then find the secret key by solving such a system. In this paper, we focus on the polynomial system solving part. The problem of solving polynomial systems over finite fields is known to be very difficult (non-deterministic polynomial-time hard complete in general). The security of many cryptographic systems is based on this problem, which makes developing algorithms for solving polynomial systems be a hot research topic in cryptanalysis.

Gröbner bases, first introduced in [6], are by now a fundamental tool for tracking this problem and become a powerful method for algebraic attacks. In addition, Gröbner bases can be used to determine optimal equations in terms of degree and/or variables in the algebraic attacks. What's more, Albrecht and Cid [14] use Gröbner bases algorithms to perform a consistency check. This allows them to determine whether given pair satisfies the considered differential characteristic. Cryptanalysis involving the Gröbner bases algorithms has been claimed to attack many cryptosystems: multivariate public key cryptosystems such as HFE [1], Minrank [2], McEliece [3], stream ciphers such as Bivium [4], hash function such as SHA-1 [5].

Finding Gröbner bases is a difficult task, which requires lots of computational resources. Algorithms to compute Gröbner bases have evolved a great deal since the first one was proposed in 1965 by Bruno Buchberger [6]. A significant leap in performance was achieved with the introduction of the F4[7] and F5[8] algorithms by Jean-Charles Faugère. In fact, F4 and F5 can be regarded as the two sides of Faugère's algorithm: F4 algorithm uses Gaussian elimination to speed up the time-consuming step of "critical pair" reductions. F5 algorithm uses a more powerful criterion to remove useless critical pairs. In recent years, many new variants of F5 are proposed and discussed, for example, EF5[9], F5C[10], G2V [11], GVW[12] and many other algorithms. However, the research of F4 is mainly focus on the implementation of algorithm. Therefore, its theory needs a further research.

A variant of F4 is designed by Antoine Joux and Vanessa Vitse[15] at SCC10 to compute the Gröbner bases of a set of polynomial systems having the similar shape. Their variant consists of the two routines F4Precomp and F4Remake. For precomputation purposes, F4Precomp run the original F4 algorithm on the first system and store only the useful polynomial multiples $(u_i, f_i)$ coming from the critical pairs. For each subsequent system, run the F4Remake to directly work on the previously stored multiples $(u_i, f_i)$. A little unfortunately, the applicability of Antoine Joux's variant is somewhat limiting. Firstly, maybe we couldn't get the Gröbner bases from F4Precomp if the initial system is difficult to solve. Secondly, polynomial systems having the similar shape doesn't mean they are the same. A little difference may lead error in the F4Remake. It is hard to guarantee the correctness of the algorithm.

In cryptanalysis, any information leakages may result in serious threat to cryptosystems. However, a serious drawback exists in the Gröbner bases based algebraic attacks, namely, we won't get any information if we couldn't work out the Gröbner bases of the polynomial equations system. In addition, lots of the cryptosystems are defined over GF(2). So computing the Gröbner bases over GF(2) is especially important. In this paper, our goals are to accelerate the F4 algorithm over GF(2) and make Gröbner bases be more practical for algebraic attacks. By using S-polynomials to replace critical pair and computing the normal form of the productions ($u_i f_i$) with respect to the field equations in certain steps, many "redundant" reductors are avoided. By slightly modifying the logic of F4 algorithm, we solve the univariate polynomials appeared in the algorithm and then back-substitute the values of the solved variables at each iteration of the algorithm. In our algorithm, even though we couldn't work out the Gröbner bases, some information of the variable still leak during the computation process. We call our variant Middle-Solving F4. We must stress that the heuristic strategy of Middle-Solving has never been applied to Gröbner bases algorithms until now. We mention that our heuristic strategy, by design, will boost the performance of all Gröbner bases algorithms in the same way as it aids F4 algorithm. We present experimental results to demonstrate that the Middle-Solving strategy has the ability to improve the F4 algorithm drastically, and make algebraic attacks be more practical.

The paper is structured as follows. First we do some preliminaries in Sect. 2. In Sect.3 we do some small improvements to make F4 be more adapted for the finite fields over GF(2). In Sect. 4 we describe our Middle-Solving F4 and introduce experimental results on various benchmark systems in Sect. 5. Sect. 6 concludes this paper.

## 2 PRELIMINARIES

This section describes the fundamental notations and the conventions in this paper. We briefly give the main definitions needed to define a Gröbner bases in a characterization useful for our purpose and simply describe the F4 algorithm.

## 2.1 Basic Knowledge

Let $K$ be a field and $R = K[x_1, x_2, \ldots, x_n]$ be the polynomial ring over the field $K$ with $n$ variables. Let $<_T$ denote a fixed admissible ordering on the monomials of $R$. The head monomial and head term of the polynomial $p \in R$ with respect to $<_T$ are denoted by $HM(p)$ and $HT(p)$ respectively. A Gröbner bases of $I = \langle F=(f_1, f_2, \ldots, f_m) \rangle$ with respect to $<_T$ is a finite list $G$ of polynomials in $I$ that satisfies the properties $\langle G \rangle = I$ and for every $p \in I$ there exists $g \in G$ satisfying $HM(g) | HM(p)$. Buchberger first found an algorithm to compute such a basis [6]. We describe Buchberger's algorithm in the following way and introduce some definitions at the same time: set $G=F$, then iterate the following two steps.

- Choose a pair $p, q \in G$ that has not yet been considered, and construct its S-polynomial $S(p,q) = \dfrac{lcm(HM(p), HM(q))}{HT(p)} \cdot p - \dfrac{lcm(HM(p), HM(q))}{HT(q)} \cdot q$

- *Top-reduce* $S(p,q)$ with respect to $G$. That is, $r_0 = S(p,q)$, and while $t = HT(r_i)$ remains divisible by $u = HT(g)$ for some $g \in G$, put $r_{i+1} := r_i - \dfrac{t}{u} \cdot g$ until no more top-reductions of $r_j$ are possible after $j$ iterations. If $r_j = 0$, we say that S($p,q$) *reduces to zero with respect to G*. if $r_j \neq 0$, we say that $S$ top-reduces to a normal form $r_j$, and append $r_j$ to $G$.

The algorithm terminates once the S-polynomials of all pairs $p, q \in G$ top-reduce to zero.

## 2.2 F4

The reduction of selected pairs is by far the biggest time-consuming part of the Buchberger's algorithm. The main idea of Faugère's F4 algorithm is to use linear algebra to simultaneously reduce a large number of pairs. F4 works with critical pairs instead of S-polynomials: the critical pair $C(f_1, f_2)$ of two polynomials $f_1$ and $f_2$ is defined as the tuple($lcm, u_1, f_1, u_2, f_2$) where $lcm$=LCM(LM($f_1$), LM($f_2$)), , the least common multiple of LM(f1) and LM(f2), and $u_i = \dfrac{lcm}{LT(f_i)}$. At each iteration step, a Macaulay-style matrix is constructed, whose columns correspond to monomials and rows to polynomials. This matrix contains the products $(u_i f_i)$ coming from the selected critical pairs (classically, all pairs with the lowest total degree $lcm$, but other selection strategies are possible) and also all polynomials involved in their reductions, which are determined during the Symbolic preprocessing phase. By computing the reduced row echelon form of this matrix, we obtain the reduced S-polynomials of all pairs considered. This algorithm, combined with an efficient implementation of linear algebra, yields very good results. A complete description of this F4 is presented below (algorithm 1, 2, 3, 4). For a more detailed discussion we refer the reader to [7].

**Algorithm 1**   F4
**Input:** $F = (f_1, f_2, \ldots, f_m) \in R^m$
**Output:** The Gröbner bases of $F$.
**Initialization:** $G := \varnothing$ and $P := \varnothing$ and $d := 0$
1. **while** $F \neq \varnothing$ **do**
2.    $f := first(F)$
3.    $F := F \setminus \{f\}$
4.    $(G, P) := Update(G, P, f)$
5. **while** $P \neq \varnothing$ **do**
6.    $d := d+1$
7.    $P_d := Select(P)$
8.    $P := P \backslash P_d$
10.   $(\tilde{F}_d^+, F_d) := Reduction(P_d, G, (F_i)_{d=1,\ldots,(d-1)})$
11.   **for** $h \in \tilde{F}_d^+$ **do**
12.     $(G, P) := Update(G, P, h)$
13. **return** $G$

**Algorithm 2**   Reduction
**Input:** $P_d$ a finite subset of selected critical pairs
   $G$ a finite subset of $R[x]$
   $\mathbb{F} = (F_k)_{k=1,\ldots,d-1}$, where $F_k$ is finite subset of $R[x]$
**Output:**   two finite subsets of $R[x]$
1.   $F :=$ Symbolic Preprocessing($P_d, G, \mathbb{F}$)
2.   $\tilde{F} :=$ Reduction to Row Echelon Form of $F$ **w.r.t.** $<$
3.   $\tilde{F}^+ := \{f \in \tilde{F} \mid HT(f) \notin HT(F)\}$
4.   **return** ($\tilde{F}^+, F$)

**Algorithm 3** Symbolic Preprocessing

**Input:** $P_d$ a finite subset of selected critical pairs
$G$ a finite subset of $R[x]$
$\mathbb{F} = (F_k)_{k=1,\ldots,d-1}$, where $F_k$ is finite subset of $R[x]$

**Output:** a finite subsets of $R[x]$
1. $F = \bigcup_{C(f_1,f_2) \in P_d} \{\text{mult}(\text{Simplify}(u_1, f_1, \mathbb{F})), \text{mult}(\text{Simplify}(u_2, f_2, \mathbb{F}))\}$
2. $Done := \text{HT}(F)$
3. **While** $\text{T}(F) \neq Done$ **do**
4.     $m$ an element of $\text{T}(F) \setminus Done$
5.     $Done := Done \cup \{m\}$
6.     **if** $m$ top reducible module $G$ **then**
7.        $m = m' * \text{HT}(f)$ for some $f \in G$ and some $m' \in T$
8.        $F := F \cup \{\text{mult}(\text{Simplify}(m', f, \mathbb{F}))\}$
9. **return** $F$

---

**Algorithm 4** Simplify

**Input:** $t \in T$ a term
$f \in R[x]$ a polynomial
$\mathbb{F} = (F_k)_{k=1,\ldots,d-1}$, where $F_k$ is finite subset of $R[x]$

**Output:** a non evaluated product, i.e. an element of $T \times R[x]$
1. **for** $u \in$ list of divisors of $t$ **do**
2.    **if** $\exists j (1 \leq j < d)$ such that $(u * f) \in F_j$ **then**
3.      $\tilde{F}_j$ is the row echelon form of $F_j$ **w.r.t.** $<$
4.      There exists a (unique) $p \in \tilde{F}_j^+$ such that $\text{HT}(p) = \text{HT}(u * f)$
5.      **if** $u \neq t$ **then**
6.        **return** Simplify$(\frac{t}{u}, p, \mathbb{F})$
7.      **else**
8.        **return** $(1, p)$
9. **return** $(t, f)$

## 3 ADAPTING F4 FOR THE FINITE FIELDS OVER GF(2)

In order to detect some hidden property of F4 over GF(2). We must adapt F4 for the finite fields over GF(2) by using some property of GF(2). And then we accelerate the speed of F4 algorithm over GF(2) without changing the framework of F4 algorithm.

### 3.1 Taking the Field Equations into Account

For cryptographic purpose, solutions in the algebraic closure are irrelevant for us. Usually, only solutions over the finite fields are of importance. For a polynomial ring $R=F_q[X]$ we can write the set of field equations of the form $x^q-x=0$ for all $x \in X$. A potential way to deal with this issue is to try to adjoin the set of field polynomials to the list of equations that we want to solve. Consequently we have to compute the Gröbner bases of $m + n$ polynomials and $n$ variables. In fact, the more equations you have the more able you are to compute a Gröbner bases. In addition, by adjoining the set of all field polynomials $F$ to the initial set of polynomials $I$, we in effect force the exponents of variables in all polynomials in the middle bases and final Gröbner bases during the computation of F4 to be fixed under the automorphism $x^q \mapsto x$. Of particular interest in the case $q=2$, where we can easily have the following simple proposition.

**Proposition 1**: By adjoining the set of field polynomials to the initial polynomial ideal in $R=F_2[X]$ with $n$ variables, the maximal degree of all polynomials in the middle bases and final Gröbner bases during the computation of F4 is at most $n$.

In fact, taking the field equations into account for Gröbner bases algorithm has also been discussed previously. We mention it here, it is just because it is a foundation of our following improvements. In addition, we want to stress its importance for F4 by some experiments and in Sec.5 we will give the real reason why adding field equations have the ability to improve the efficiency of the algorithm. Experimental results to compare Field-equations-adding F4 (FE-F4 for short) with the original F4 for some HFE systems and some classical benchmarks (Cyclic 6, Gonnet83) are presented in Table 1. The total time in seconds and the max memory in Megabytes are represented by "Time" and "Max-mem" respectively. From the experimental results, we can see adding field equations has the ability to improve the F4 algorithm drastically.

**Table 1.** Performance of FE-F4 versus original F4 while computing some benchmarks

| Test case | Time | | Max-mem | |
|---|---|---|---|---|
| | F4 | FE-F4 | F4 | FE-F4 |
| HFE(17,5) | 2.465 | 0.047 | 4.832 | 3.099 |
| HFE(17,6) | 15.054 | 0.094 | 12.812 | 3.132 |
| HFE(17,7) | 97.578 | 0.188 | 77.944 | 3.282 |
| Gonnet83 | 14.259 | 1.684 | 19.380 | 5.988 |
| Cyclic 6 | 2.247 | 0.172 | 3.608 | 2.818 |

## 3.2 Critical Pair or S-polynomial?

In a sense, S-polynomial is equivalent to critical pair. However, after taking the field equations into account and doing some special processing during the F4 algorithm over GF(2), there exists a greatly different impact on process of Reduction between the using of S-polynomials and critical pairs.

F4 works with critical pairs. The productions $(u_i f_i)$ coming from the selected critical pairs are put into the set $F$ during each iteration step of F4. And then subroutine *Symbolic Preprocessing* is executed to select and calculate all eligible reductors of $F$. All reductors are also put into $F$ and then the reduced S-polynomials are calculated by doing Gaussian elimination to the Macaulay-style matrix constructed from $F$. In order to better understand the following improvement, here we must explain the generating process of reductors. A reductor $h$ generated from middle bases $G$ that satisfies the properties $\text{LT}(h) = m' * HT(f) \in T(F) \backslash HT(F)$ for some $m' \in T(F) \backslash HT(F)$ and some $f \in G$, then $f' = \text{mult}(\text{Simplify}(m', f, \mathbb{F})) = \tilde{m}' * \tilde{f}$ is put into $F$. Repeat this process until no reductors generate. We can clearly see that the number of monomials in the all productions $(u_i f_i) \in F$ coming from the selected critical pairs directly impact the number of generated reductors and the final size of matrix. So our goal is to reduce the number of the monomials in $F$. The exponents of all variables in the polynomials in middle bases are fixed under 2 by adjoining the set of all field polynomials to input system $I$ over GF(2). This leads many the same monomials exist in the two productions $(u_i f_i)$, $(u_j f_j)$ coming from a critical pair $(lcm, u_i, f_i, u_j, f_j)$. Unfortunately, these same monomials will mislead us to find some "redundant" reductors. For example: $I = < f_1 = xy + x, f_2 = yz + z + 1, f_3 = xz + 1, f_4 = x + y + z >$, we assume the critical pair $C(f_1, f_2) = (xyz, z, f_1, x, f_2)$ is selected. Then $zf_1 = xyz + xz$ and $xf_2 = xyz + xz + x$ are put into set $F$. Then the algorithm tries to generate the reductors whose leading terms are $xz$ and $x$ during the first iteration of Symbolic Preprocessing phase. The goal of subprogram *Reduction* is to obtain the reduced S-polynomials of all critical pairs considered. If we compute the S-polynomial first: $S(f_1, f_2) = zf_1 - xf_2 = xyz + xz - (xyz + xz + x) = x$ and put the S-polynomial into set $F$, we only need to generate the reductor whose leading term is $x$ during the first iteration step of Symbolic Preprocessing; nay, we could avoid many redundant computations in the next iteration of Symbolic Preprocessing because the "redundant" reductors whose leading term *xz* may generate new monomials. So, instead of working with critical pairs, we suggest the F4 algorithm works with S-polynomials over GF(2).

In order to ensure the correctness of the algorithm, We should slightly modify some subprograms of F4 if we want to work with S-polynomials instead of critical pairs over GF(2). The set *NEWF* should be added in the subprogram Symbolic Preprocessing to record all the reductors. The set $\tilde{F}^+$ in subprogram Reduction is selected from polynomials in $\tilde{F}$ satisfies $HT(f) \notin HT(NEWF)$ instead of $HT(f) \in HT(F)$. This is because our goal is to obtain the reduced S-polynomials after row echelon computing phase. $HT(NEWF)$ is the set of all leading terms of reductors. A polynomial *f* satisfies $HT(f) \notin HT(NEWF)$ in the row echelon form of the matrix is a reduced S-polynomial.

By observing the polynomials during the computation process of F4 algorithm, we found that although the exponents of all variables in the polynomials in middle bases are fixed under 2 by adjoining the set of all field polynomials to the set of polynomials *I* over GF(2), there are still some polynomials with variables' exponents higher than 2 appear during the computation process. This is because multiplications are done in the productions ($u_i f_i$) coming from the selected critical pairs and reductors. In order to get a greater efficiency of our improvement, it is better to compute the normal form of the productions with respect to field equations by top-reducing the productions ($u_i f_i$) appear with respect to field polynomials. Thus the exponents of all variables in the productions are fixed under 2. More monomials will be eliminated after constructing the S-polynomials from the critical pairs and less new monomials are introduced into matrix from the reductors. For example, Let $I = <f_1=xy+yz, f_2=xz+yz+1, f_3=xy+x>$, we assume the selected critical pair is $C(f_1, f_2)$, then $S(f_1, f_2) = zf_1 - yf_2 = xyz + yz^2 - (xyz + y^2z + y) = yz^2 - y^2z - y$. But if we top-reduce the productions $zf_1$ and $yf_2$ with respect to field polynomials first, $S(f_1, f_2)' = NormalForm(zf_1) - NormalForm(yf_2) = xyz + yz - (xyz + yz + y) = -y$. We need only to consider monomial *y* instead of $yz^2, y^2z, y$ in the next computation.

Algorithm 5 and Algorithm 6 are the modified Reduction and Symbolic Preprocessing respectively. For convenience, we name the improvement S-F4.

**Algorithm 5** Reduction

**Input:** $P_d$ a finite subset of selected critical pairs
         $G$ a finite subset of $R[x]$
         $\mathbb{F} = (F_k)_{k=1,\ldots,d-1}$, where $F_k$ is finite subset of $R[x]$

**Output:** two finite subsets of $R[x]$

1. $F$, $NewF := $ Symbolic Preprocessing$(P_d, G, \mathbb{F})$
2. $\tilde{F} := $ Reduction to Row Echelon Form of $F$ **w.r.t.** $<$
3. $\tilde{F}^+ := \{f \in \tilde{F} \mid HT(f) \notin HT(NEWF)\}$
4. **return** $(\tilde{F}^+, F)$

---

**Algorithm 6** Symbolic Preprocessing

**Input:** $P_d$ a finite subset of selected critical pairs
         $G$ a finite subset of $R[x]$
         $\mathbb{F} = (F_k)_{k=1,\ldots,d-1}$, where $F_k$ is finite subset of $R[x]$

**Output:** a finite subsets of $R[x]$

1. $F = \bigcup_{C(f_1, f_2) \in P_d} \{NF(mult(Simplify(u_1, f_1, \mathbb{F}))) - NF(mult(Simplify(u_2, f_2, \mathbb{F})))\}$
2. $Done := HT(F)$, $NewF := \{\}$
3. **While** $T(F) \neq Done$ **do**
4.    $m$ an element of $T(F) \setminus Done$
5.    $Done := Done \cup \{m\}$
6.    **if** $m$ top reducible module $G$ **then**
7.    $m = m' * HT(f)$ for some $f \in G$ and some $m' \in T$
8.    $F := F \cup \{NF(mult(Simplify(m', f, \mathbb{F})))\}$
9.    $NewF := NewF \cup \{NF(mult(Simplify(m', f, \mathbb{F})))\}$
10. **return** $F$, $NewF$

Remark. "$NF$" is the abbreviation for "Normal Form"

## 4 MIDDLE-SOLVING F4

In this section, we will slightly modify the logic of the F4 algorithm to make it be more practical for cryptanalysis (especially algebraic attacks) to solve cryptosystems over GF(2).

     Usually, the key of algebraic attacks is solving multivariate polynomial equations. Gröbner bases is one of the most efficient algorithms to solve the system of equations. More precisely, we can decompose the algorithm in two distinct steps when we use Gröbner bases to solve the system of equations: first we should compute the Gröbner bases of the initial polynomials system, then we get the value of the variables from the Gröbner bases with other algorithm (e.g. with Berlekamp's algorithm). Gröbner bases

just serves as an intermediate step to solve systems of polynomial equations symbolically. In the suitable term orders (e.g. lexicographical orders), the solutions of the Gröbner bases can be easily computed by successively eliminating variables. However, a serious drawback exists in the Gröbner bases based algebraic attacks, namely, we won't get any information if we couldn't work out the Gröbner bases of the polynomial equations system. This drawback greatly restricts the practicability of algebraic attacks. Our goal is to overcome this drawback, getting some information during the computation process of F4 algorithm and accelerating the F4 algorithm.

In many experiments with our improved F4 algorithm described Section.3, we have observed that after the Gaussian elimination step certain univariate polynomials appear. They are just treated like any other polynomials in the F4 algorithm. We must stress here that our ultimate aim is to obtain the solution of the initial equations system, not the Gröbner bases of the initial equations system. If those univariate polynomials have only one solution, they could be used beneficially in the algorithms. We show that they should deserve a special treatment. For the practical purposes, we slightly modify the logic of the F4 algorithm. At each iteration of the algorithm, solve these univariate polynomials and back-substitute the values of the solved variables to all polynomials in the set of middle bases $G$, the set of critical pairs $P$, and the set $\mathbb{F}$. And then delete the zero elements in these sets. We can image that the benefits of such improvement is huge. It will cause a series of chain reactions, this makes the whole algorithm get relatively simpler than before, even though only a value of one variant is obtained. And even though we couldn't work out the Gröbner bases, some information of the variable still leak during the computation process. We name our improvement Middle-Solving F4. We must stress that the heuristic strategy of Middle-Solving has never been applied to Gröbner bases algorithms until now. We mention that our heuristic strategy, by design, will boost the performance of all Gröbner bases algorithms in the same way as it aids F4 algorithm.

Solving the univariate polynomials is a basic idea for solving equations. Existing algorithms for Gröbner bases are generally only concerned with how to quickly get Gröbner bases. The strategy of Middle-Solving breaks the consistent thinking of Gröbner basis algorithms for solving equations. Middle-Solving F4 introduces no spurious solutions, destroys no solutions. In fact, during the computation process of Middle-Solving F4, once a value of variable is obtained, the following operation is equivalent to computing Gröbner bases with respect to the unsolved variables. During the process of solving a system of multivariate polynomial equations, the less number of equations and variables, the less computational complexity needed to find the solution. In the worst case that no one value of the variable is obtained, the Middle-Solving F4 is just equivalent to F4. A formal description and a flowchart of the Middle-Solving F4 are presented in Algorithm 7 and Figure 1 respectively.

**Algorithm 7** Middle-Solving F4

**Input:** $F = (f_1, f_2, \ldots, f_m) \in R^m$
**Output:** The Gröbner bases of $F$.
**Initialization:** $G := \emptyset$ and $P := \emptyset$ and $d := 0$
1. **while** $F \neq \emptyset$ **do**
2.    $f := first(F)$
3.    $F := F \setminus \{f\}$
4.    $(G,P) := Update(G,P,f)$
5. **while** $P \neq \emptyset$ **do**
6.    $d := d+1$
7.    $P_d := Select(P)$
8.    $P := P \setminus P_d$
9.    $(\tilde{F}_d^+, F_d) := Reduction(P_d, G_{d-1}, F_d)$
10.    $UP := [f : f \text{ in } \tilde{F}_d^+ \mid \text{IsUnivariate}(f)]$;
11.    **if** $UP$ is not Empty **then**
12.      $R := [f : f \text{ in } UP \mid \#Roots(\text{UnivariatePolynomial}(f)) \text{ eq } 1]$;
13.      **for** $r$ in $R$ **do**
14.        $root := Solve(\text{UnivariatePolynomial}(r))$;
15.      $Renew(G, P, \mathbb{F})$    // Back-substitute the values of the solved variables
16.    **for** $h \in \tilde{F}_d^+$ **do**
17.      $(G,P) := Update(G,P,h)$
18. **return** $G$

**Figure 1.** A general description of Middle-Solving F4

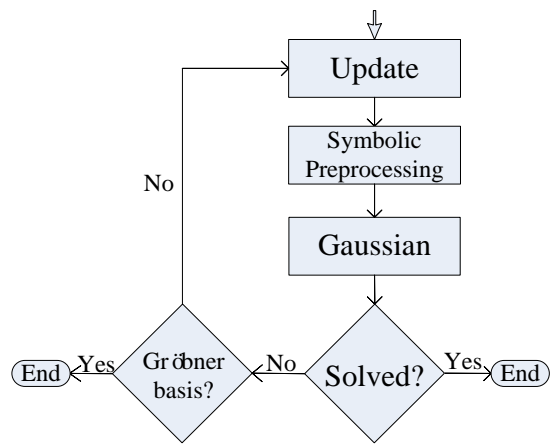

## 5 EXPERIMENTAL RESULTS AND ANALYSIS

In this section, experimental results are presented to compare our improvement to Faugère's F4. We take an interest in solving systems of some classical benchmarks (Cyclic 6 , Gonnet83) and some HFE(17,$n$) cryptosystems generated by Segers [13], where 17 is the degree restriction of the hidden polynomial used to generate the public key polynomials, $n$ is number of variables and generated polynomials. The implementations in this section are all written by Magma version (V2.11-11). Our F4 program is written according to the pseudo code of Faugère's F4 and not join other implementation techniques, such as sparse matrix techniques, etc. We regard it as a benchmark for our improvement to prove the effectiveness of our improvement. So our improvements can be applied to the implementation of the different people, including the Gröbner bases program building in Magma

In order to comprehend why our variant can improve the F4 drastically, timings and memory alone are not enough for the comparison. They are only the intuitive reflection to the efficiency of the algorithm and depend heavily on the efficiency of some hidden algorithms. It is well-known that the most time-consuming part of F4 consists in the reduction operations. Thus an accurate comparison would consider:

- The number of critical pairs considered, which is denoted by "C-Pair".
- The largest number of rows in the reduction matrix, which is denoted by "L-Matrix".
- The total number of reductors considered, which is denoted by "Reductor".
- The highest degree of final Gröbner bases, which is denoted by "H-Deg-GB".
- The number of final Gröbner bases, which is denoted by "#GB".

Table 2 gives some key datas during the computation process of the experiments we did in Table 1. From Table 2, we can easily get the reason that why adding field equations has the ability to improve the F4 algorithm drastically over GF(2). After adding the field equations to the initial equations system, the number of critical pairs considered and the largest number of rows in the reduction matrix decrease drastically. This is mainly because that lots of the redundant critical pairs are avoided by the Gebauer–Möller Criterion during the update phase. And the limitation of the exponents of the variables also plays a very important role. The datas need to be computed decease a lot, so it is not surprising that the time and the max memory decrease as described in Table 1. We notice that the highest degree of final Gröbner bases decrease a lot. It is simpler for us to solve the Gröbner bases by other algorithm.

Adding field equations to the initial equations system can accelerate the F4 algorithm and needs less memory. So we regard the FE-F4 as a benchmark. The following experiment results of our variants all compare to the FE-F4 unless stated.

Table 2. Internal data in both original F4 and FE-F4 while computing some benchmarks

|             | C-Pair |       | L-Matrix |       | H-Deg-GB |       |
|-------------|--------|-------|----------|-------|----------|-------|
| Test case   | F4     | FE-F4 | F4       | FE-F4 | F4       | FE-F4 |
| HFE(17,5)   | 98     | 32    | 311      | 49    | 32       | 3     |
| HFE(17,6)   | 209    | 51    | 795      | 77    | 17       | 3     |
| HFE(17,7)   | 401    | 82    | 2053     | 106   | 15       | 3     |
| Gonnet83    | 1154   | 309   | 1216     | 571   | 6        | 4     |
| Cyclic 6    | 180    | 60    | 206      | 75    | 13       | 6     |

Table 3 shows the results after we work with the S-polynomials instead of critical pairs for HFE systems. "$n$" denotes the number of equations and the number of variables for each initial system. The total time of Reduction subprogram is denoted by "R-Time". In each case, FE-F4 and S-F4 compute almost the same number of critical pairs, because working with the S-polynomials instead of critical pairs almost only affect the number of reductors during the Reduction phase. In Faugère's F4, the number of reductors generated by Symbolic Preprocessing and the final size of matrix are decided by number of the monomials in selected critical pairs. However, in our S-F4, they are decided by number of the monomials in S-polynomials generated from selected critical pairs. It is obviously that the number of the monomials in S-polynomials is far less than the monomials in critical pairs over GF(2). In addition, the number of the monomials farther decreases by top-reducing the productions appear with respect to field equations during the Reduction phase. Due to the decrease of the number of reductors generated by Symbolic Preprocessing and the final size of matrix, the workload of Symbolic Preprocessing and Gaussian elimination in the subprogram Reduction decreases. So the total time of subprogram Reduction of S-F4 is far less than FE-F4.

Table 3. Reductions performed by the FE-F4 and S-F4 over GF(2) for HFE(17,$n$)

|     | C-Pair |      | L-Matrix |      | H-Deg-GB |      | Reductor |      | R-Time  |        |
|-----|--------|------|----------|------|----------|------|----------|------|---------|--------|
| $n$ | FE-F4  | S-F4 | FE-F4    | S-F4 | FE-F4    | S-F4 | FE-F4    | S-F4 | FE-F4   | S-F4   |
| 9   | 243    | 243  | 785      | 208  | 6        | 6    | 3924     | 879  | 6.972   | 1.216  |
| 10  | 455    | 455  | 1463     | 310  | 7        | 7    | 10623    | 2090 | 31.608  | 4.591  |
| 11  | 705    | 711  | 2368     | 527  | 8        | 7    | 22395    | 3670 | 127.635 | 13.85  |
| 12  | 794    | 794  | 1650     | 537  | 7        | 7    | 10664    | 2678 | 40.932  | 9.921  |
| 13  | 1407   | 1407 | 2773     | 960  | 7        | 7    | 22090    | 5276 | 181.396 | 42.446 |

*Remark.* The Reduction subprogram of S-F4 here includes the computation of S-polynomials and the computation the normal form to the productions.

Tables 4 and 5 show the internal data and the performance of the Middle-Solving F4 for HFE systems respectively. The number of solved variables and total iteration round of the algorithm are represented by "#Solved" and "Round" respectively. Through the improvements in the Section 3, at least four univariate equations appear in our experiments (HFE(17,10) and HFE(17,11)). Once a variable is obtained, the following computation may be far easier. By comparing the internal data, including the number of reductors, the max size of matrix and total iteration round, we can see the workload of Middle-Solving F4 is far less than FE-F4. This is the reason why Middle-Solving F4 always outperforms EF-F4 in terms of memory and time. The more difficult the equations system is, the more obvious our advantage will be. Since the values of some variables are obtained, the highest degree and the number of final Gröbner bases computed by Middle-Solving F4 are both lower than FE-F4.

**Table 4.** Internal data in both FE-F4 and Middle-Solving F4 for HFE(17,*n*)

|   | L-Matrix | | #Solved | Round | | Reductor | |
|---|---|---|---|---|---|---|---|
| *n* | FE-F4 | M-S | M-S | FE-F4 | M-S | FE-F4 | M-S |
| 9  | 785  | 153 | 5 | 13 | 12 | 3924  | 617 |
| 10 | 1463 | 215 | 4 | 20 | 18 | 10623 | 1357 |
| 11 | 2368 | 274 | 4 | 26 | 25 | 22395 | 2573 |
| 12 | 1650 | 504 | 7 | 17 | 14 | 10664 | 1785 |
| 13 | 2773 | 588 | 7 | 21 | 18 | 22090 | 3697 |

**Table 5.** Performance of Middle-Solving F4 versus FE-F4 for HFE(17,*n*)

|   | #GB | | H-Deg-GB | | Time | | Max-mem | |
|---|---|---|---|---|---|---|---|---|
| *n* | FE-F4 | M-S | FE-F4 | M-S | FE-F4 | M-S | FE-F4 | M-S |
| 9  | 148 | 76  | 6 | 3 | 7.971   | 1.747  | 11.571  | 4.049 |
| 10 | 272 | 226 | 7 | 4 | 36.676  | 8.128  | 33.434  | 5.115 |
| 11 | 385 | 343 | 8 | 4 | 138.544 | 21.06  | 79.745  | 7.278 |
| 12 | 457 | 219 | 7 | 3 | 58.984  | 20.311 | 36.056  | 8.658 |
| 13 | 769 | 478 | 7 | 3 | 254.5   | 95.862 | 101.219 | 16.028 |

## 6 CONCLUSION AND FUTRUE WORK

For practical purpose, we in-depth research the F4 algorithm for cryptanalysis (especially algebraic attacks) to solve cryptosystems over GF(2). In order to overcome the serious drawback of the Gröbner bases based algebraic attacks that no information leak if we couldn't work out the Gröbner bases of the polynomial equations system, a new variant named Middle-Solving F4 is presented in this paper. Even though we couldn't work out the Gröbner bases, some information of the variable still leak during the computation process. So our variant is well adapted for algebraic attacks on cryptosystems. Experimentally, Middle-Solving F4 outperforms Faugère's F4 in terms of memory and time. We believe that the heuristic strategy of Middle-Solving is a general approach that can improve most of the Gröbner bases algorithms. And the Middle-Solving F5 is in the process of being implemented and should be available soon.

Guessing some bits during the computing Gröbner bases in many cases is a surprisingly effective way of accelerating the solution of the polynomial systems of equations. It is more effective but intractable than Hybrid approach [16], which guess some bits before computing Gröbner bases. We plan to mix between the "guess and determine" method and our Middle-Solving F4 or other Gröbner bases algorithms to farther improve the practicability of algebraic cryptanalysis.